# Development of an embedded-atom method potential of Ni-Mo alloys for electrocatalysis / surface compositional studies

Ambesh Gupta[1], Chinmay Dahale[1], Soumyadipta Maiti[1], Sriram Goverapet Srinivasan[2], Beena Rai[1]

[1]TCS Research, Tata Consultancy Services Limited, Plot No. 2 & 3, MIDC-SEZ, Rajiv Gandhi Infotech Park, Hinjewadi Phase III Pune, 411057, Maharashtra, India

[2]Tata Consultancy Services Limited, IIT-Madras Research Park, Block A, Second Floor, Phase - 2, Kanagam Road, Taramani, Chennai 600113, Tamil Nadu, India

Email:   soumya.maiti@tcs.com, s.goverapet@tcs.com

## Abstract

Ni-Mo superalloys have emerged as materials of choice for a diverse array of applications owing to their superior mechanical properties, exceptional corrosion and oxidation resistance, electrocatalytic behavior, and surface stability. Understanding and optimizing the surface composition of Ni-Mo alloys is critical for enhancing their performance in practical applications. Traditional experimental surface analysis techniques, while informative, are often prohibitive in terms of cost and time. Likewise, theoretical approaches such as first-principle calculations demand substantial computational resources and it is difficult to simulate large structures. This study introduces an alternative approach utilizing hybrid Monte-Carlo / Molecular Dynamics (MC/MD) simulations to investigate the surface composition of Ni-Mo alloys. We report the development of an optimized Embedded-Atom Method (EAM) potential specifically for Ni-Mo alloys, carefully parameterized using empirical lattice constants and formation energies of elemental and face-centered cubic (FCC) Ni-Mo solid solution alloys. The reliability of the EAM potential is corroborated via the evaluation of equations of state, with a particular focus on reproducing structural properties. Utilizing this validated potential, MC/MD simulations were performed to understand the depth-wise variations in the compositions of Ni-Mo alloy nanoparticles and extended surfaces. These simulations reveal a preferential segregation of nickel on surface, and molybdenum in sub-surface layer. Due to this preferential segregation, it is imperative to consider surface segregation while tailoring the surface properties for targeted applications.

## 1. Introduction

Traditionally Ni-based alloys and superalloys have been widely used in advanced aircraft engines for turbine blades due to their improved high temperature mechanical properties, oxidation resistance [1] etc. Recently developed Ni-based superalloys incorporate various alloying elements like Mo, Cr, Ta, Re, W and Zr [2],[3],[4]. Among the alloying elements, molybdenum has been observed as a key alloying element for the development of nickel-based superalloys [2],[5]. Ni-Mo alloy subsystem exhibited exceptional mechanical strength, resistance to thermal creep deformation, good surface stability, and resistance to corrosion and oxidation [6],[7],[8],[9], making them very useful for a multitude of applications in aerospace and chemical industries. Furthermore, Ni-Mo alloys were found to be superior materials for electrocatalytic processes, where their excellent catalytic activity and stability were utilized in hydrogen production, fuel cells, and other energy conversion systems [10],[11],[12],[13]. This wide range of utilities underscores the importance of Ni-Mo alloys in modern technology and engineering.

Catalysis being a 'surface-governed' phenomena, a detailed understanding of the composition and atomic structure at various facets of a material is essential to derive structure property relations and explain or predict the origins of their superior performance (or otherwise). Moreover, the surface characteristics also affect the mechanical properties, durability, and interactions with other materials.

Previous studies have shown varied results for the Ni-Mo system regarding their surface composition. Using Low-Energy Electron Diffraction (LEED), Auger Electron Spectroscopy (AES) and Electron Spectroscopy for Chemical Analysis (ESCA) measurements, Marcus *et al.* [14] observed that in Ni-Mo systems with 2 and 6 atomic % Mo content, the surface was enriched in molybdenum. Similarly, using ESCA measurements, Highfield *et al.* [15] found an Ni/Mo ratio of ~6 and ~17 for $Ni_9Mo$ and $Ni_{24}Mo$ alloys, respectively, indicating an enrichment in Mo content at the surface. In yet another study, Martinez *et al.* [11] measured a surface composition of 22% Mo for a Ni-15%Mo alloy via Energy Dispersive X-ray Spectroscopy (EDX) experiments. While these results point to the enrichment of Mo at the surface, it is yet unclear if such enrichment is due to the preferential segregation of Mo at the surface or the preferential etching of Ni in sputtering.

In fact, theoretical investigations in past show that nickel enriched surface is more stable than a random or molybdenum enriched surface in Ni$_3$Mo alloy [16] [17]. Recently, another study [18] reported that Ni-Mo alloy systems containing Mo atoms in the sub-surface layer were thermodynamically stable. Clearly, a detailed study of the surface composition and atomic structures in different Ni-Mo alloys and their facets would not only help in understanding the segregation trends, but also create realistic surface models for further evaluation of these alloys as catalysts in various chemical conversions.

Experimental investigations are limited by the difficulty in characterizing the detailed atomic arrangement of elements in these alloys as a function of depth away from the surface layer. In addition, preparation of suitable samples for surface characterization experiments is complex in itself. As noted earlier, detailed knowledge of the surface atomic structure is essential to gauge the activity of an alloy catalyst since subsurface atoms are known to modulate the catalytic activity at the surface. In such a situation, molecular simulations using classical interatomic potentials are immensely useful as they can provide detailed atomic scale surface structures across various alloy composition and surface cleavage planes. While numerous embedded atom method potentials have been developed for Ni and Mo in combination with various elements [19],[20],[21],[22],[23],[24], lack of an accurate interatomic potential to model Ni-Mo interactions has been a bottleneck in the investigation of surface structures and compositions of Ni-Mo alloys. A few of the earlier reported potentials for Ni-Mo systems include the EAM potential by Zhou formalism [24], Finnis-Sinclair (F-S) potential for Ni-Zr-Mo ternary system [25], Spectral Neighbor Analysis Potential (SNAP) using Machine Learning (ML) [26] and Modified Analysis Embedded Atom Method (MAEAM) potential for studying irradiation effects on Ni-Mo alloys [27]. Other than the SNAP potential, none of the others consider substitutional solid solutions. Moreover, none of these potentials are able to capture the thermodynamics of solid solutions accurately and are as such unsuitable for identifying thermodynamically stable surface structure(s) and composition(s). To overcome this limitation, in this study, we have developed an EAM potential for Ni-Mo alloy system which accurately captures the thermodynamics of both pure elements and various Ni$_x$Mo solid solutions (x=2,3,4,8). We have further used this potential in hybrid MC/MD simulations to investigate the preferential surface segregation and the atomic scale structure in both spherical nanoparticles and extended surfaces of various Ni$_x$Mo alloys. Our results indicate that Ni atoms preferentially segregate to the surface in both spherical nanoparticles and extended surfaces of all stable Ni$_x$Mo alloys. Details of the ab initio calculations used to generate reference data are presented in section 2, further the development of the EAM potential and hybrid MC/MD simulations are discussed. Section 3 presents the validation of the developed EAM potential and results from hybrid MC/MD simulations. Our results and discussion on developed EAM and its possible applications are summarized in Section 4.

## 2. Computational Methods and Details

### a. DFT calculations

Density Functional Theory (DFT) calculations were carried out using the Vienna Abinitio Simulation Package (VASP) [28],[29] to compute the formation energies of Ni$_x$Mo alloys (x=2,3,4,8). The electronic states were expanded on a planewave basis up to a kinetic energy cutoff of 500 eV. The Brillouin zone was sampled using a Γ-centered (6x6x6) ***k***-point mesh [30]. The Perdew-Burke-Ernzerhof (PBE) functional [31] was used to describe the electron exchange-correlation interactions while the core-valence interactions were represented using the Projector Augmented Wave (PAW) approach [32]. The alloy structures were modelled as a 2x2x2 FCC supercell with 32 atoms in them. The initial lattice parameters for the alloys were computed with Zen's law [33] using the lattice parameters of the pure elements, namely FCC Ni and BCC Mo. First, a geometry optimization was carried out at this lattice parameter. Subsequently, volume relaxation with constant atomic fractional coordinates and cell shape (i.e., ISIF = 7 in VASP) was carried out. These calculations were deemed to have converged when the force on each atom fell below 2 meV/Å. The electronic SCF convergence criteria in each geometry optimization step were set to 10$^{-6}$ eV. The lattice parameters of the pure metals were obtained by optimizing their conventional unit cells. The formation energies (eV/atom) of the alloys ($\Delta f_{Ni_xMo}$) are calculated using the following equation:

$$\Delta f_{Ni_xMo} = E_{Ni_xMo} - xE_{Ni} - E_{Mo} \quad \quad \dots (1)$$

where $E_{Ni_xMo}$ is the total energy of the alloy and $E_{Ni}$ and $E_{Mo}$ are the total energies of pure Ni (FCC) and pure Mo (BCC) at their ground states. Since the alloys are considered to be solid solutions, 13 unique random atomic configurations were generated for each composition (Ni$_x$Mo), and its formation energy was calculated as an average of these 13 different configurations.

To validate the EAM potential, the single point energies of the most stable structure for each alloy composition were also calculated at volumetric strains of 1%, 2%, 5% and 10% in expansion and compression.

### b. Construction of EAM potential

**b.1. EAM potential formalism**

EAM is a many-body interatomic potential which represents the total energy of a system as the sum of a pair-wise interaction energy and an embedding energy.

$$E_{total} = \sum_i F_i \left( \sum_{j \neq i} f_j(r_{ij}) \right) + \frac{1}{2} \sum_{i,j\,(i \neq j)} \varphi_{ij}(r_{ij}) \quad \ldots (2)$$

Here, $F_i$ is the embedding energy function of atom type $i$, $\varphi_{ij}$ is the pair-wise interaction energy function between atoms $i$ and $j$, $f$ is a spherically symmetric electron density function that accounts for contribution from atom $j$ to the electronic charge density at site $i$.

We have used embedding and electronic density functions for Ni and Mo available from Zhou database as they were shown to provide accurate elemental properties [24]. Also, all the EAM potential related functions were already scaled in the same range, so there was no need for re-scaling. Like-pair interaction energy of the elements was calculated using equation 3 [24].

$$\varphi(r) = \frac{A \exp\left[-\alpha \left(\frac{r}{r_e} - 1\right)\right]}{1 + \left(\frac{r}{r_e} - \kappa\right)^{20}} - \frac{B \exp\left[-\beta \left(\frac{r}{r_e} - 1\right)\right]}{1 + \left(\frac{r}{r_e} - \lambda\right)^{20}} \quad \ldots (3)$$

Here, $\varphi$ is the pair interaction energy, $r_e$ is the equilibrium spacing between the nearest neighbors, A, B, $\alpha, and\ \beta$ are the four adjustable parameters, and $\kappa\ and\ \lambda$ are two additional parameters for the cutoff. These parameters for the Ni and Mo elements are as shown in Table 1 [24].

Table 1: Parameters for the pair interaction energy of Ni and Mo elements [24].

| Parameter | Ni | Mo |
|---|---|---|
| A | 0.429046 | 0.708787 |
| B | 0.633531 | 1.120373 |
| $r_e$ | 2.488746 | 2.728100 |
| $\alpha$ | 8.383453 | 8.393531 |
| $\beta$ | 4.471175 | 4.476550 |
| $\kappa$ | 0.443599 | 0.176977 |
| $\lambda$ | 0.820658 | 0.353954 |

The hetero pair-interaction energy between Ni and Mo ($\varphi_{NiMo}$) atoms was modeled using equation 4 [34]

$$\varphi_{NiMo}(r) = b\, \varphi_{Ni}(r+c) + d\, \varphi_{Mo}(r+e) \qquad \ldots (4)$$

where, $\varphi_{Ni}$ is like-pair interaction energy function of Ni and $\varphi_{Mo}$ is like-pair interaction energy function of Mo, and $b, c, d, e$ are four parameters to be fit.

**b.2. EAM potential fitting**

A Particle-Swarm Optimization (PSO) algorithm [35] was used to fit the parameters of the EAM potential. As we took the elemental functions from prior literature, there was no need to optimize for them. Thus, only the parameters of the hetero-pair function (i.e., b, c, d and e in equation (4)) were optimized. The squared error cost function was used for optimization, as shown in equation 5.

$$Cost = \sum_i w_{f_i} * \left(\Delta f_{i,EAM} - \Delta f_{i,DFT}\right)^2 + \sum_i w_{a_i} * \left(a_{i,EAM} - a_{i,DFT}\right)^2 \qquad \ldots (5)$$

Here, $\Delta f_{i,EAM}$ and $\Delta f_{i,DFT}$ were the EAM and DFT computed formation energies of structure *i*, while $a_{i,EAM}$ and $a_{i,DFT}$ were their respective lattice parameters. $w$ was a weight fraction of the two properties in the cost function. The PSO algorithm used a swarm size of 40, acceleration coefficients ($c_1$ and $c_2$) as 0.5, Inertia weight ($\omega$) as 0.5 and the maximum number of iterations were set to 200. The convergence criteria were set to default as $1 \times 10^{-8}$ and the weight fractions of the properties were tuned occasionally in the range of (0-10). The parameters search range was set to (-5,5). All the EAM simulations were performed using the LAMMPS software [36].

### c. Hybrid Monte-Carlo / Molecular Dynamics Simulations

Hybrid Monte Carlo Molecular Dynamics (MC/MD) simulations were carried out to study the preferential segregation of Ni / Mo at the surfaces of various Ni$_x$Mo alloys [37],[38]. Five different systems were considered in these simulations – a spherical and cubic nanoparticle and extended surfaces of (111), (110) and (100) miller indices. Firstly, a large cubic FCC structure containing about 6000 atoms was constructed for each alloy composition following which they were energy minimized while allowing for both the lattice parameters and the atomic coordinates to change. Next, MD simulations were carried out for 50ps in the NPT ensemble at a target temperature of 1200K and 1 bar pressure using an MD timestep of 1 fs. A Nose-Hoover barostat [39],[40],[41] was used to maintain the system temperature and pressure at the desired value. The value of the lattice parameter at the target temperature was then taken as an average over the last 15ps of simulation. These parameters were used to generate the structure of the nanoparticles as well as extended surfaces, each containing about 6000 atoms.

Subsequently, the structures of these systems were evolved in MC/MD simulations. In these simulations, the positions of two different atoms were chosen randomly and swapped. Following the swap, the atomic coordinates were relaxed using the conjugate gradient method. Acceptance of the swap was decided based on the Metropolis criterion [42], as outlined below.

$$Swap = \begin{cases} Accepted \,;\, \Delta U < 0 \\ Accepted\ with\ probability\,;\, 0 < p < e^{-\frac{\Delta U}{k_B T}} \\ Rejected \,;\, p > e^{-\frac{\Delta U}{k_B T}} \end{cases}$$

where, $\Delta U$ is the difference in potential energy of system before and after swap of the atom in system, $k_B$ is the Boltzmann's constant, $T$ is the temperature of the system and $p$ is a random number between 0 and 1.

During the MC/MD simulations, all the atoms were allowed to swap in the nanoparticles while only the top four to six atomic layers were considered for atom swaps in extended surfaces. For the slab structures, the top four to six layers of around 0.8 nm of the system correspond to the surface. Remaining layers correspond to the bulk and the atoms were not allowed to swap in this region. Surface evolution was performed at 1200 K until 10 swaps per atom for all the system. Following system evolution, layer-wise and radial composition profiles were obtained for the extended surfaces and nanoparticles, respectively. Five different random initial configurations were considered for each system and the results were obtained as an average over these configurations. Visualizations of alloy's atomic structures were generated using the OVITO Software [43].

## 3. Results and Discussion

### a. Fitting the EAM potential

The parameters of the cross-pair interaction energy were fitted against DFT computed values of the lattice parameters and formation energies of various Ni-Mo alloys. The optimized parameter values are given in table 2. Figure 1 shows that the pair interaction energy vs interatomic distance plot, computed using these parameters, is smooth and continuous.

Table 2 : Fitted parameters of cross-pair interaction energy function for Ni and Mo.

| Parameter | Optimized Value |
|---|---|
| b | 1.7705228 |
| c | -0.055409 |
| d | -0.2133725 |
| e | -0.0059651 |

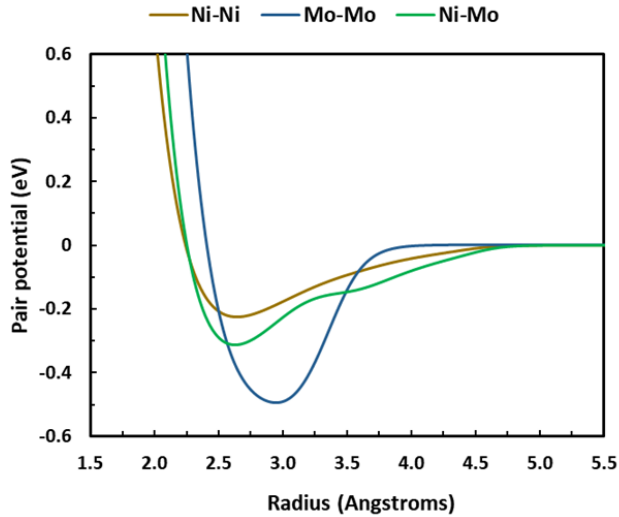

*Figure 1 : Pair potential as function of radius for Ni-Ni, Mo-Mo, and Ni-Mo.*

The formation energies and lattice parameters of various $Ni_xMo$ alloys computed with DFT and the optimized EAM potential are given in table 3 while figure 2 shows a comparison of the formation energies with prior published EAM potential of Zhou *et al.* [24] and a SNAP potential [26]. Clearly, our EAM potential closely reproduces DFT values with a mean error of only 5.7 meV/atom in formation energies and 1.94% in lattice parameters. In comparison, Zhou *et al.'s* [24] EAM potential had an error of 169 meV/atom in formation energies and 0.39% in lattice parameters. These values for the SNAP potential [26] were 67 meV/atom and 3.24%. Furthermore, both the earlier potentials gave (incorrectly) a positive formation energy for various $Ni_xMo$ phases, indicating the absence of any stable Ni-Mo bimetallic alloy. In contrast, the potential developed in this work correctly captures the trends in the formation energies of $Ni_xMo$ alloys. To further validate the developed potential, equation of states (EOS) for $Ni_xMo$ alloys were computed and compared against their DFT values. Figure 3 clearly shows that the potential is closely able to reproduce the DFT computed EOS.

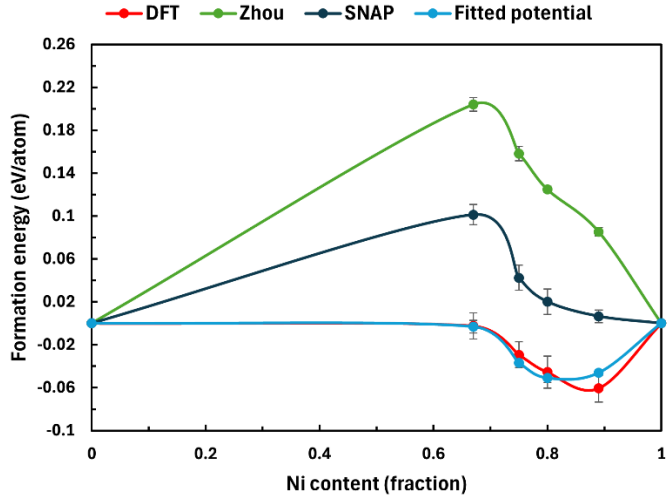

*Figure 2 : Formation energy of alloy at different Ni content from different techniques (DFT, MD using Zhou [24] , SNAP [26] and fitted potentials).*

*Table 3 : Lattice parameter, and formation energy of the system using DFT and newly constructed potential from MD.*

| Alloy | Methodology | Lattice parameter ± Std. dev (Å) | Formation energy ± Std. dev (eV/atom) |
|---|---|---|---|
| Ni (FCC) | DFT | 3.5070 | 0 |
| | EAM | 3.5204 | 0 |
| $Ni_2Mo$ (FCC) | DFT | 3.6846 ± 0.0033 | -0.0024 ± 0.0123 |
| | EAM | 3.5892 ± 0.0071 | -0.0032 ± 0.0060 |
| $Ni_3Mo$ (FCC) | DFT | 3.6374 ± 0.0032 | -0.0293 ± 0.0120 |
| | EAM | 3.5575 ± 0.0068 | -0.0369 ± 0.0021 |
| $Ni_4Mo$ (FCC) | DFT | 3.6063 ± 0.0028 | -0.0454 ± 0.0150 |
| | EAM | 3.5451 ± 0.0036 | -0.0510 ± 0.0042 |
| $Ni_8Mo$ (FCC) | DFT | 3.5726 ± 0.0018 | -0.0606 ± 0.0129 |
| | EAM | 3.5314 ± 0.0021 | -0.0462 ± 0.0024 |
| Mo (BCC) | DFT | 3.15 | 0 |
| | EAM | 3.15 | 0 |

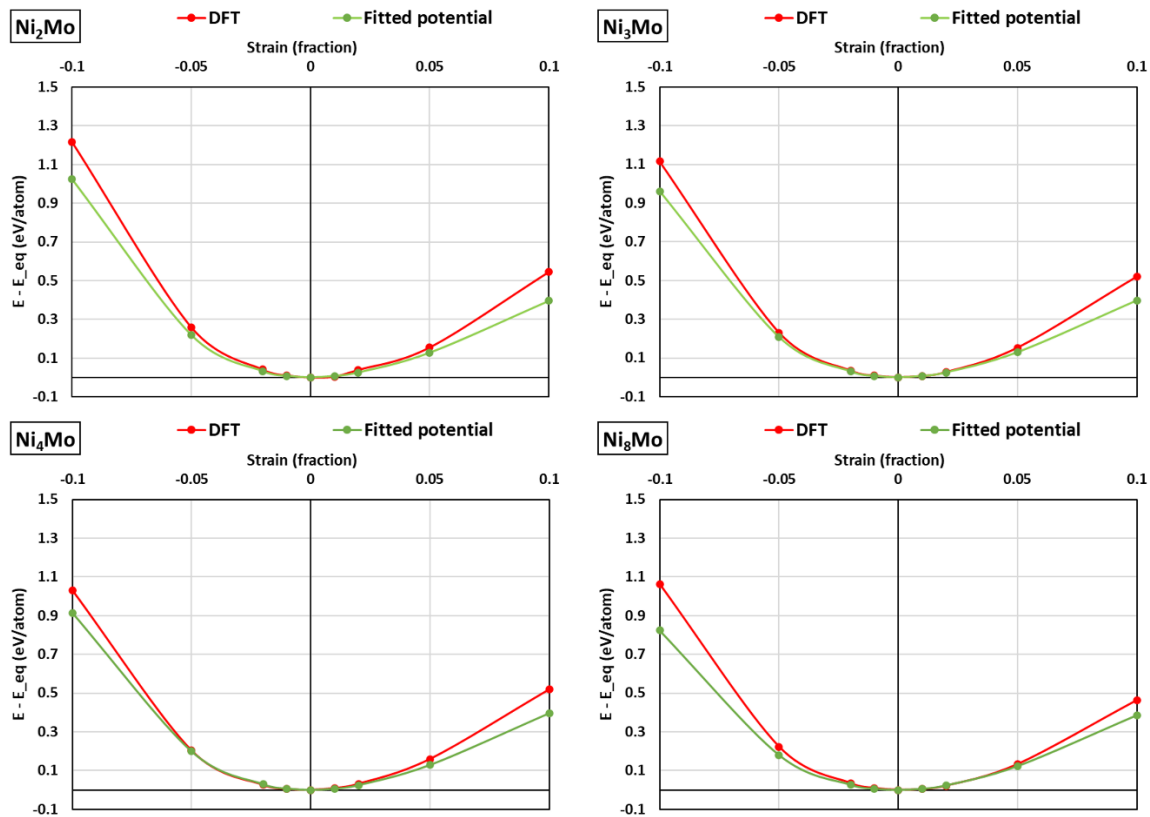

*Figure 3 : Comparison of equation of states of $Ni_2Mo$, $Ni_3Mo$, $Ni_4Mo$, and $Ni_8Mo$ under strain of -0.1 to +0.1 in all directions from DFT and MD using fitted potential.*

### b. Surface Segregation in $Ni_xMo$ alloys

The optimized EAM potential was used to study the preferential segregation in nanoparticles and extended surfaces of various $Ni_xMo$ alloys via MC/MD simulations. Although the MC/MD simulations were carried out till 10 swaps per atom, figure S1 of supplementary information shows that the energy and compositions of the systems converged within 5 atomic swaps.

**b.1. Segregation in spherical and cubic nanoparticles (NPs)**

We studied the segregation in spherical NPs with a radius of 2.57 nm in all the $Ni_xMo$ (x = 2,3,4,8) compositions, containing around 6000 atoms. To begin with, Ni and Mo atoms in the NPs were distributed randomly and MC swaps were allowed between all atoms. Unlike planar surfaces that are periodic in nature, the NPs are non-periodic in nature. As a result, each atom at the surface of NPs may witness a different chemical neighborhood as well as different coordination numbers.

Figure 4 shows the structure of the initial spherical NP formed by random distribution of atoms and the evolved structure after the MC/MD simulations for Ni₃Mo composition.

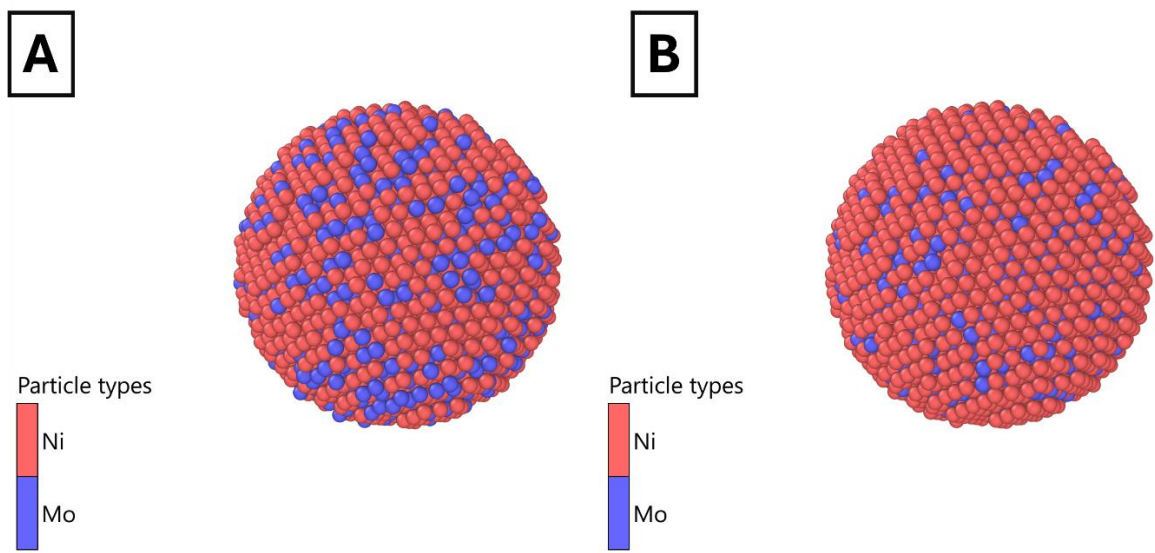

*Figure 4: Snapshots of Ni₃Mo spherical NP (a) Initial random configuration, (b) segregated structure after the MC/MD evolution.*

A significant amount of atom migration and segregation is observed in the spherical NPs. The radial compositional profile in figure 5 shows that Ni atoms have segregated to the surface of the NP, while sub-surface of the NP is dominated by the Mo atoms. Error bars in the figure at various radius points denote the standard deviation over the five unique configurations taken for the calculations.

This segregation of nickel in surface layer can be attributed to the low surface energies of FCC nickel (2.21 J/m$^2$) as compared to FCC molybdenum (2.62 J/m$^2$) [44]. Mixing enthalpy of the solid solution was observed to be negative using Miedema model, which implies a no significant segregation [45],[46],[47]. The computed mixing enthalpies for alloys based on Miedema model and atomic strain are presented in table S1 of supplementary information. The segregation energy of molybdenum impurity in a nickel FCC matrix was observed to be +0.18 eV/atom suggesting a moderate anti segregation of molybdenum in the lattice. Thus, addition of Mo in Ni lattice prefers to remain in the interior of the lattice [48].

Interestingly, all the compositional variations are within the first three to four layers including surface, which corresponds to max depth of around 0.6 nm. Beyond this region, the composition of the NP remains the same as of the bulk composition. Similar variations were observed in other alloy compositions as well ($Ni_2Mo$, $Ni_4Mo$, and $Ni_8Mo$), for which the depth wise compositional variations are shown in figure S2 of supplementary information.

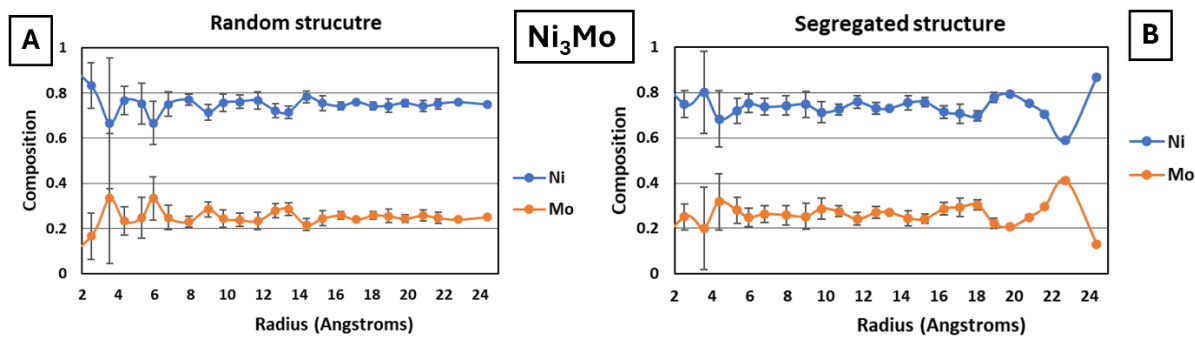

*Figure 5: Variation in composition of $Ni_3Mo$ nanoparticle as a function of radius (a) Initial random configuration (b) Segregated structure. At highest radius which is surface layer, average nickel composition increases while in 2nd layer it decreases as compared to initial configuration.*

To further explore the effect of NP shape on element segregation trends, cubic shaped NPs with an initial edge length of 4 nm, containing approximately 6600 atoms were studied. Unlike spherical NP, each face of the cubical NP belongs to the same family of surface orientations (i.e., {100} type). Thus, the physical environment of atoms at surface is similar, except for atoms at the edges or corners of the NP. However, this was not the case in spherical NPs.

The average surface and sub-surface composition over six faces of NP structures at different compositions of alloy were averaged over five unique configurations. Figure 6 shows that the trends in compositional variation in surface and sub-surface layers are similar to that seen in spherical NPs. Surface is dominated by the nickel atoms while molybdenum atoms have segregated in the sub-surface layer.

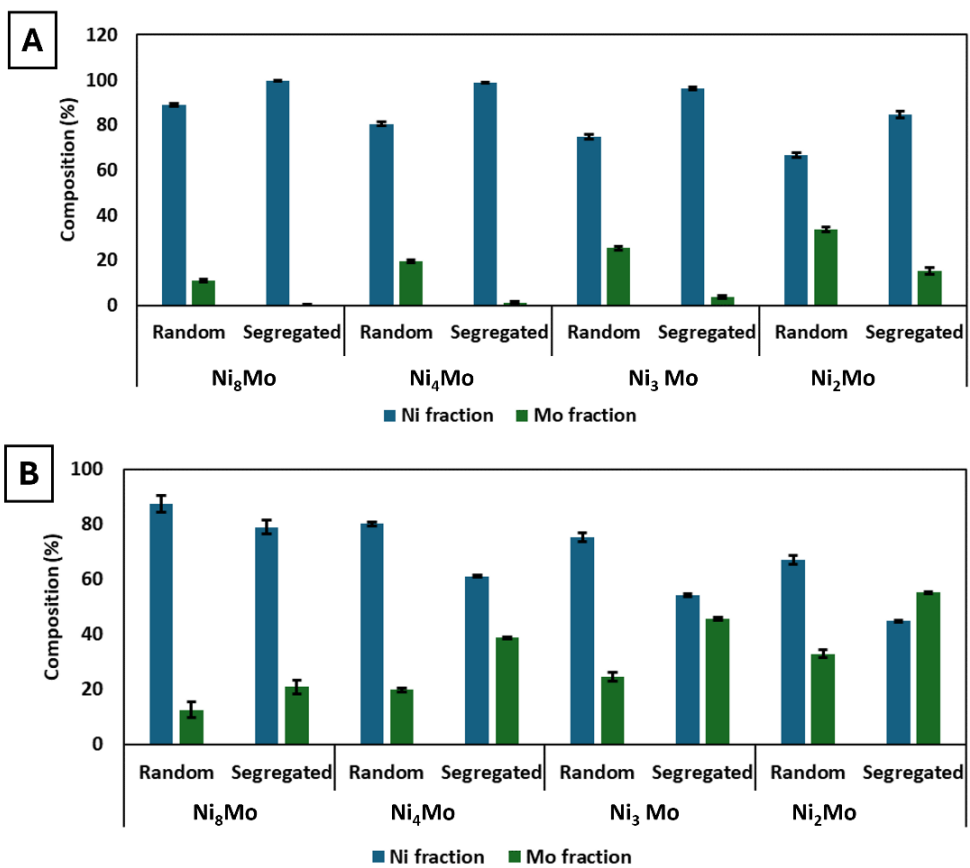

*Figure 6: Average composition of alloy structures (standard deviation as error bar) of initial and final configurations in (A) Surface layer, (B) Sub-surface layer. Nickel segregates in surface layer and Mo segregates in sub-surface layer.*

For all the nanoparticles, Ni segregation is observed to be more prominent on surface layers in cubical NPs as compared to the spherical NPs of same composition. The Ni composition in random and segregated states of surface and sub-surface layer in NPs is shown in Table 4. The extent of segregation of nickel in surface layer is almost similar irrespective of the NP's composition. However, the increase in concentration of Mo in sub-surface layer is positively correlated with the overall Mo content in the alloy structure. For instance, $Ni_8Mo$ cubical NP shows an ~ 8% increase in Mo concentration, while $Ni_2Mo$ cubical NP shows an increase of ~ 23% in the sub-surface layer.

Table 4: Averaged nickel composition in surface and sub-surface layers of cubical and spherical nanoparticles of $Ni_xMo$ alloy structure (Random and Segregated). Standard deviation is computed across the five unique atomic structures of the nanoparticles of each $Ni_*Mo$ alloy (x = 2,3,4,8).

| Shape | Layer | Structure | $Ni_2Mo$ | $Ni_3Mo$ | $Ni_4Mo$ | $Ni_8Mo$ |
|---|---|---|---|---|---|---|
| Spherical | Surface | Random | 65.21 ± 0.64 | 74.87 ± 0.73 | 79.85 ± 0.37 | 88.03 ± 0.57 |
| | | Segregated | 74.11 ± 3.88 | 86.93 ± 0.86 | 87.77 ± 1.31 | 96.44 ± 1.81 |
| | Sub-surface | Random | 67.06 ± 0.73 | 75.94 ± 0.62 | 80.33 ± 0.94 | 89.13 ± 0.93 |
| | | Segregated | 52.41 ± 4.12 | 58.84 ± 1.04 | 68.86 ± 3.48 | 83.34 ± 0.89 |
| Cubical | Surface | Random | 66.47 ± 1.10 | 74.68 ± 1.04 | 80.54 ± 0.76 | 88.98 ± 0.66 |
| | | Segregated | 84.75 ± 1.57 | 96.30 ± 1.04 | 98.77 ± 0.40 | 99.76 ± 0.13 |
| | Sub-surface | Random | 67.06 ± 1.48 | 75.42 ± 1.65 | 80.19 ± 0.68 | 87.46 ± 2.88 |
| | | Segregated | 44.79 ± 0.46 | 54.36 ± 0.60 | 61.19 ± 0.41 | 79.05 ± 2.43 |

**b.2. Segregation in flat slab surfaces**

After simulations on NPs, segregation in the slab surfaces is studied. MC/MD simulations are carried out specifically on the (111), (100), and (110) low index surfaces. NPs simulations suggested that composition generally varied within 0.8 nm region beneath the surface. Thus, for MC/MD simulation in surfaces, we allowed swaps between atoms only in top four layers for (100) and (111) surfaces and top 6 layers for (110) surface. After every swap, all atoms are relaxed using Conjugated Gradient (CG) energy minimization scheme. Surface composition of all the solid solutions for various surfaces is shown in Table 5. The layer-wise composition profiles of $Ni_3Mo$ for the initial random and evolved/segregated structures for all types of surface slabs are shown in figure 7 while composition profiles for the other alloys are show in figures S3 – S5 of the supporting information.

*Table 5: Averaged nickel composition in surface layer of different types of surface slabs of $Ni_xMo$ alloy structure (Random and Segregated). Standard deviation is computed across the 5 unique arrangements of atoms in the slabs of each $Ni_xMo$ alloy (x = 2,3,4,8).*

| Alloy | Surface | Structure | Composition | Alloy | Surface | Structure | Composition |
|---|---|---|---|---|---|---|---|
| **$Ni_2Mo$** | **(100)** | Random | 67.76 ± 2.96 | **$Ni_3Mo$** | **(100)** | Random | 76.12 ±2.91 |
| | | Segregated | 90.56 ±1.02 | | | Segregated | 98.11 ± 1.12 |
| | **(110)** | Random | 67.92 ± 4.19 | | **(110)** | Random | 74.44 ± 3.76 |
| | | Segregated | 97.78 ± 0.84 | | | Segregated | 99.44 ± 0.31 |
| | **(111)** | Random | 67.96 ± 2.23 | | **(111)** | Random | 75.93 ± 15.95 |
| | | Segregated | 84.95 ± 1.64 | | | Segregated | 95.69 ±0.89 |
| **$Ni_4Mo$** | **(100)** | Random | 79.64 ± 1.11 | **$Ni_8Mo$** | **(100)** | Random | 88.67 ±1.83 |
| | | Segregated | 99.03 ± 0.33 | | | Segregated | 99.39 ± 0.53 |
| | **(110)** | Random | 80.42 ± 1.73 | | **(110)** | Random | 88.68 ± 1.90 |
| | | Segregated | 99.79 ± 0.31 | | | Segregated | 99.93 ± 0.16 |
| | **(111)** | Random | 79.03 ± 1.46 | | **(111)** | Random | 89.54 ± 1.40 |
| | | Segregated | 96.67 ± 1.03 | | | Segregated | 99.63 ± 0.31 |

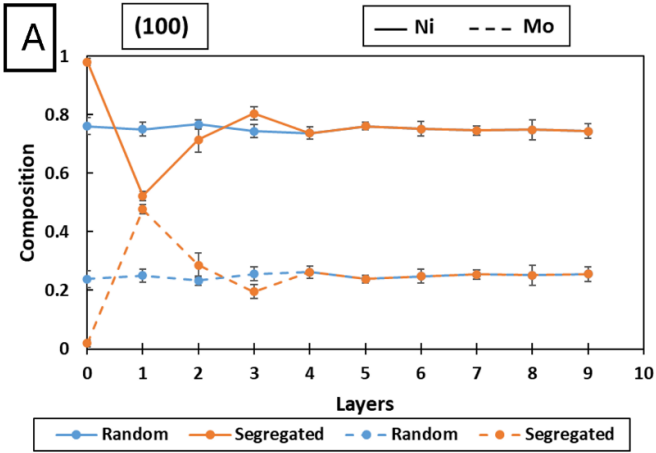

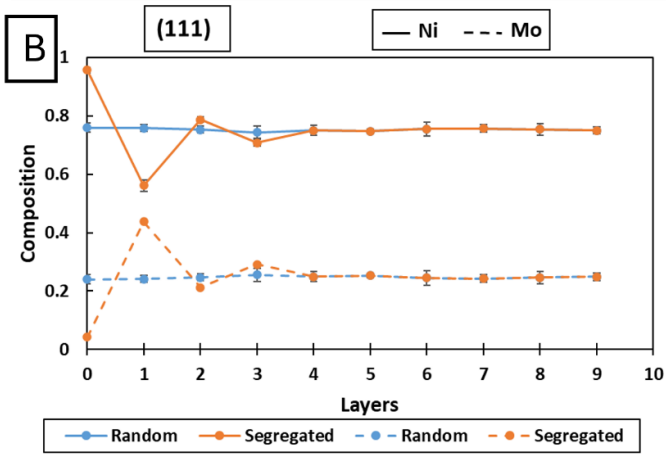

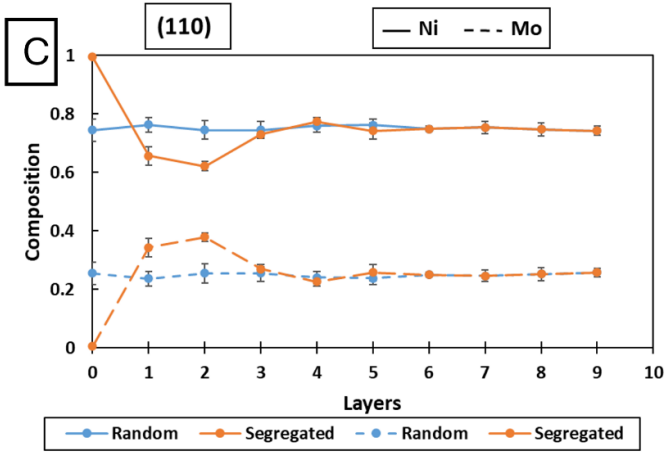

Figure 7: Variation of the average nickel and molybdenum composition as a function of layers in the random and segregated structures of Ni$_3$Mo for (100), (110), and (111) surfaces.

Evolved structures in Figure 7 show that the Ni atoms segregate in the surface layer for all the three surface orientations while the sub-surface layer is enriched in Mo atoms. However, for the other layers, there is not much of a compositional change as compared to random structures except for the (110) surface. On this surface, we observed the 2nd sub-surface layer to be the most depleted in Ni content. The extent of Ni segregation among low index surfaces follows the order: (111) < (100) < (110). In addition, the extent of Mo enrichment in the sub-surface layer increases with an increase in the Mo content of the alloy, as shown in figure 8.

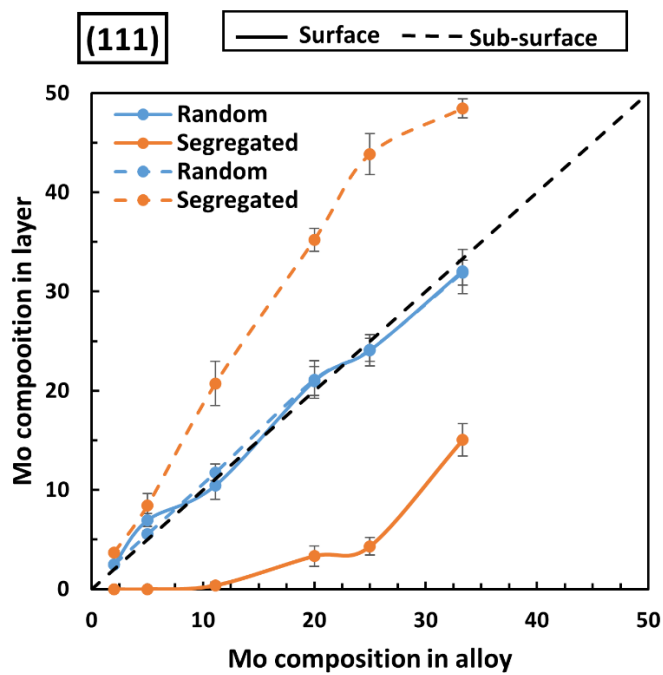

*Figure 8: Composition of molybdenum (Mo) in surface and sub-surface layers as a function of molybdenum composition (atomic %) in alloy for (111) surface slabs.*

The preferential segregation of element(s) at an alloy surface is driven by a combination of two factors – surface energy and the mixing enthalpy. Table S1 of the supporting information reports the mixing enthalpy of the alloys calculated using the Miedema model. Table S2 presents the surface energy values computed with the EAM potential from this work and those reported from an earlier DFT study [44]. From Table S2, we notice that the surface energy of all the low index facets of FCC Ni are lower than those of Mo. Thus, although the mixing energy is favorable between Ni and Mo at all compositions, it appears that the surface energy is the dominant factor in the observed segregation trends. These results are in agreement with earlier reports that showed Ni enrichment in the (111), (100) and (201) surfaces of Ni$_3$Mo alloy [16],[17].Furthermore, prior calculations also indicated a high segregation energy of Mo in Ni FCC lattice [48] and point to Mo presence in sub-surface layer as a thermodynamically stable configuration [18]. Such preferential segregation has a profound impact on the catalytic activity of Ni$_x$Mo alloys. DFT calculations have hypothesized that the presence of Mo atoms in the sub-surface layer weakens the hydrogen binding energy at the surface Ni atoms, thereby enhancing the activity of Ni$_x$Mo alloys for hydrogen evolution reaction [18]. Thus, a proper estimate of the surface and sub-surface compositions in Ni$_x$Mo alloys, as facilitated by the EAM potential developed in this work, will enable a more realistic screening of these alloy catalysts for various chemical conversions. Finally, the rapid equilibration of surface compositions in our MC/MD simulations suggest that the segregation kinetics in Ni$_x$Mo alloys could be fast, with sub-surface Mo atom effects being prevalent across multiple applications.

## 4. Conclusion

In this study, a n-body EAM potential of Ni-Mo has been developed successfully. Developed EAM Potential achieves consistently low RMSEs values for lattice parameter formation energies for the FCC Ni-Mo binary alloy systems. The formation energy of solid solutions from the developed EAM are compared with earlier available EAM, SNAP potentials, and have proved to be much better. The developed potential predicts the relative energy under compressive and tensile equation of states accurately. Thus, the developed potential can capture not only the thermodynamic properties, but also mechanical and structural properties of $Ni_xMo$ systems.

Using the developed potential, MC/MD simulations were carried out to investigate surface segregation in $Ni_xMo$ alloy systems. Nanoparticles and surfaces are found to be enriched in nickel atoms while sub-surface is enriched with molybdenum atoms irrespective of alloy composition. Difference of surface energies of individual elements could be the driving force of surface segregation. The extent of segregation in the different type of surface slabs is estimated with highest in (110) type of planes while (111) surface shows the least extent of Ni segregation. The effects of surface segregation in Ni-Mo alloys system should be considered while calculating the activity and selectivity for various catalytic conversions as surface segregation might alter the adsorbate binding strength at the surface.

**AUTHOR CONTRIBUTIONS**

S.M, A.G, C.D and S.G.S jointly conceived the idea and the implementation plan. A.G and C.D carried out the calculations and their analysis. S.M, A.G, C.D and S.G.S jointly wrote the manuscript. All authors discussed the results, their implications and commented on the manuscript.

**ACKNOWLEDGEMENTS**

This research was fully supported by the TCS-CTO organization. We thank the TCS-RnI Infra team for providing necessary support for the execution of this work.

**CONFLICTS OF INTERESTS**

There are no conflicts of interest to declare.

# References

[1] Reed, Roger C. *The superalloys: fundamentals and applications*. Cambridge university press, 2008.

[2] W. Zhao, W. Li, Z. Sun, S. Gong and L. Vitos, "Tuning the plasticity of Ni-Mo solid solutions in Ni-based superalloys by ab initio calculations," *Materials & design,* vol. 124, pp. 100-107, 2017. doi: 10.1016/j.matdes.2017.03.057

[3] D. Cai, L. Xiong, W. Liu, G. Dun and M. Yao, "Development of proccessing maps for a Ni-based superalloy," *Materials Characterization,* vol. 58, no. 10, pp. 941-946, 2007. doi: 10.1016/j.matchar.2006.09.004

[4] Y. Gao, J. Dong, L. Wang, W. Liu, J. Zhang and L. Lou, "Enhancing microstructural stability and creep properties by Ta addition in Ni-based single crystal superalloys," *Materials Science and Engineering: A,* vol. 909, p. 146873, 2024. doi: 10.1016/j.msea.2024.146873

[5] C. Y., X. Zhao, W. Xia, Q. Yue, Y. Gu, X. Wei, H. Bei, Y. Dang and Z. Zhang, "Effect of Mo on microstructural stability of a 4th generation Ni-based single crystal superalloy," *Journal of Materials Research and Technology,* vol. 21, pp. 2672-2681, 2022. doi: 10.1016/j.jmrt.2022.10.072

[6] J. Hu, Y. N. Shi, X. Sauvage, G. Sha and K. Lu, "Grain boundary stability governs hardening and softening in extemely fine nanograined metals," *Science,* vol. 355, no. 6331, pp. 1292-1296, 2017. doi: 10.1126/science.aal5166

[7] X. Li, S. Xu, Q. Zhang, S. Liu and J. Shuai, "Complex strengthening mechanism in nanocrystalline Ni-Mo alloys revealed by a machine-learning interatomic potential," *Journal of Alloys and Compounds,* vol. 952, p. 169964, 2023. doi: 10.1016/j.jallcom.2023.169964

[8] H. Tawancy, "Comparative corrosion behavior of Ni-Mo and Ni-Mo-Cr alloy for applications in reducing environments," *Journal of Materials Science,* vol. 41, pp. 8359-8362, 2006. doi: 10.1007/s10853-006-0990-y

[9] J. Wang, L. Han, X. Li, Y. Huang, Y. Liu and Z. Wang, "Temperature-dependent evolution of strength of nanocrystalline Ni(Mo) alloys at the Mo solubility limit," *Materials Science and Engineering: A,* vol. 786, p. 139326, 2020. doi: 10.1016/j.msea.2020.139326

[10] N. V. Krstajić, V. D. Jović, L. Gajić-Krstajić, B. M. Jović, A. L. Antozzi and G. N. Martelli, "Electrodeposition og Ni-Mo alloy coatings and their characterization as cathodes for hydrogen evolution in sodium hydroxides solution," *International Journal of Hydrogen energy,* vol. 33, no. 14, pp. 3676-3687, 2008. doi: 10.1016/j.ijhydene.2008.04.039

[11] S. Martinez, M. Metikoš-Huković and L. Valek, "Electrocatalytic properties of electrodeposited Ni–15Mo cathodes for the HER in acid solutions: Synergistic electronic effects," *Journal of Molecular Catalysis A: Chemical,* vol. 245, no. 1-2, pp. 114-121, 2006. doi: 10.1016/j.molcata.2005.09.040

[12] J. R. McKone, B. F. Sadtler, C. A. Werlang, N. S. Lewis and H. B. & Gray, "Ni–Mo nanopowders for efficient electrochemical hydrogen evolution," *ACS Catalysis,* vol. 3, no. 2, pp. 166-169, 2013. doi: 10.1021/cs300691m

[13] F. Bao, E. Kemppainen, I. Dorbandt, R. Bors, F. Xi, R. Schlatmann, R. v. d. Krol and S. Calnan, "Understanding the hydrogen evolution reaction kinetics of electrodeposited nickel-molybdenum in acidic, near-neutral, and alkaline conditions," *ChemElectroChem,* vol. 8, no. 1, pp. 195-208, 2021. doi: 10.1002/celc.202001436

[14] P. Marcus, C. Chandler, H. Chadli and P. Wynblatt, "Surface composition and structure of nickel-molybdenum single crystal alloys," *Applied surface science,* vol. 37, no. 1, pp. 33-43, 1989. doi: 10.1016/0169-4332(89)90971-9

[15] J. Highfield, E. Claude and K. Oguro, "Electrocatalytic synergism in Ni/Mo cathodes for hydrogen evolution in acid medium: a new model," *Electrochimica Acta,* vol. 44, pp. 2805-2814, 1999. doi: 10.1016/S0013-4686(98)00403-4

[16] Y. Yu, J. Zhang, W. Xiao, J. Wang and L. Wang, "First-principles study of surface segregation in bimetallic $Ni_3M$ (M= Mo, Co, Fe) alloys with chemisorbed atomic oxygen," *physica status solidi (b),* vol. 254, no. 6, p. 1600810, 2017. doi: 10.1002/pssb.201600810

[17] J. Wijten, R. Riemersma, J. Gauthier, L. Mandemaker, M. Verhoeven, J. Hofmann, K. Chan and B. Weckhuysen, "Electrolyte effects on the stability of Ni– Mo cathodes for the hydrogen evolution reaction," *ChemSusChem,* vol. 12, no. 15, pp. 3491-3500, 2019. doi: 10.1002/cssc.201900617

[18] R. Patil, M. Kaur, S. House, L. Kavalsky, K. Hu, S. Zhong, D. Krishnamurthy, V. Viswanathan, J. Yang, Y. Yan and J. Lattimer, "Reversible alkaline hydrogen evolution and oxidation reactions using Ni–Mo catalysts supported on carbon," *Energy Advances,* vol. 2, no. 9, pp. 1500-1511, 2023. doi: 10.1039/D3YA00140G

[19] Y. Mishin, F. D and M. Mehl, "Interatomic potentials for monoatomic metals from experimental data and ab initio calculations," *Physical Review B,* vol. 59, no. 5, p. 3393, 1999. doi: 10.1103/PhysRevB.59.3393

[20] Y. Mishin, "Atomistic modeling of the γ and γ′-phases of the Ni–Al system," *Acta Materialia,* vol. 52, no. 6, pp. 1451-1467, 2004. doi: 10.1016/j.actamat.2003.11.026

[21] G. Bonny, N. Castin and T. D., "Interatomic potential for studying ageing under irradiation in stainless steels: the FeNiCr model alloy," *Modelling and Simulation in Materials Science and Engineering,* vol. 21, no. 8, p. 085004, 2013. doi: 10.1088/0965-0393/21/8/085004

[22] A. Voter and S. Chen, "In High temperature ordered intermetallics alloys," in *MRS Symposia Proceedinngs*, 1987. doi: 10.1557/PROC-81-3

[23] D. Smirnova, K. A.Y. and S. S.V., "A ternary EAM interatomic potential for U-Mo alloys with xenon," *Modelling and Simulation in Materials Science and Engineering,* vol. 21, no. 3, p. 035011, 2013. doi: 10.1088/0965-0393/21/3/035011

[24] X. Zhou, R. Johnson and H. Wadley, "Misfit-energy-increasing dislocations in vapor-deposited CoFe/NiFe multilayers," *Physical Review B,* vol. 69, no. 14, p. 144113, 2004. doi: 10.1103/PhysRevB.69.144113

[25] M. Yang, S. Li and Y. Li, "Atomistic modeling to optimize composition and characterize structure of Ni-Zr-Mo metallic glasses," *Physical Chemistry Chemical Physics,* vol. 17, no. 20, pp. 13355-13365, 2015. doi: 10.1039/C5CP00512D

[26] X.-G. Li, C. Hu, C. Chen, Z. Deng, J. Luo and S. P. Ong, "Quantum-accurate spectral neighbor analysis potential models for Ni-Mo binary alloys and fcc metals," *Physical Review B,* vol. 98, no. 9, p. 094104, 2018. doi: 10.1103/PhysRevB.98.094104

[27] L. Lang, K. Yang, Z. Tian, H. Deng, F. Gao, W. Hu and Y. Mo, "Development of a Ni-Mo interatomic potential for irradiation simulation," *Modelling and Simulation in Materials Science and Engineering,* vol. 27, no. 4, p. 045009, 2019. doi: 10.1088/1361-651X/ab1407

[28] G. Kresse and J. Furthmüller, "Efficiency of ab initio total energy calculations for metals and semiconductors using a plane-wave basis set," *Computational materials science,* vol. 6, no. 1, pp. 15-50, 1996. doi: 10.1016/0927-0256(96)00008-0

[29] G. Kresse and J. Furthmüller, "Efficient iterative schemes for ab initio total-energy calculations using a plane-wave basis set," *Physical Review B,* vol. 54, no. 16, p. 11169, 1996. doi: 10.1103/PhysRevB.54.11169

[30] H. J. Monkhorst and J. D. Pack, "Special points for Brillouin-zone integrations," *Physical Review B,* vol. 13, no. 12, p. 5188, 1976. doi: 10.1103/PhysRevB.13.5188

[31] G. Kresse and D. Joubert, "From ultrasoft pseudopotentials to the projector augmented-wave method," *Physical Review B,* vol. 59, no. 3, p. 1758, 1999. doi: 10.1103/PhysRevB.59.1758

[32] J. P. Perdew, K. Burke and M. & Ernzerhof, "Generalized gradient approximation made simple," *Physical review letters,* vol. 77, no. 18, p. 3865, 1996. doi: 10.1103/PhysRevLett.77.3865

[33] E. Zen, "Validity of "Vegard's law"," *American Mineralogist ,* Vols. 523-524, pp. 5-6, 1956.

[34] Y. Shen, L. Liu, S. Mi, H. Gong and S. Zhou, "Construction of a n-body Fe-Cu potential and its application in atomistic modeling of Fe-Cu solid solutions," *Journal of Applied Physics,* vol. 127, no. 4, p. 045104, 2020. doi: 10.1063/1.5129015

[35] L. Miranda, "PySwarms: a research toolkit for Particle Swarm Optimization in Python," *Journal of Open Source Software,* vol. 3, no. 21, p. 433, 2018. doi: 10.21105/joss.00433

[36] S. Plimpton, "Fast parallel algorithms for short-range molecular dynamics," *Journal of computational physics,* vol. 117, no. 1, pp. 1-19, 1995. doi: 10.1006/jcph.1995.1039

[37] S. Mishra, S. Maiti, B. Dwadasi and B. Rai, "Realistic microstructure evolution of complex Ta-Nb-Hf-Zr high-entropy alloys by simulation techniques," *Scientific reports,* vol. 9, no. 1, p. 16337, 2019. doi: 10.1038/s41598-019-52170-0

[38] C. Dahale, S. Srinivasan, M. S., S. Maiti and B. Rai, "Surface segregation in the AgAuCuPdPt high entropy alloy: insights from molecular simulations," *Molecular Systems Design & Engineering,* vol. 7, no. 8, pp. 878-888, 2022. doi: 10.1039/D2ME00045H

[39] S. Nose, "A molecular dynamics method for simulations in the canonical ensemble," *Molecular Physics,* vol. 52, no. 2, pp. 255-268, 1984. doi: 10.1080/00268978400101201

[40] W. Hoover, "Canonical dynamics: Equilibrium phase-space distributions," *Physical Review A,* vol. 31, no. 3, pp. 1695-1697, 1985. doi: 10.1103/PhysRevA.31.1695

[41] G. T. D. K. Martyna, "Constant pressure molecular dynamics algorithms," *The Journal of Chemical Physics,* vol. 101, pp. 4177-4189, 1994. doi: 10.1063/1.467468

[42] W. Hastings, "Monte Carlo Sampling Methods Using Markov Chains and Their Applications," *Biometrika,* vol. 57, no. 1, pp. 97-109, 1970. doi: 10.2307/2334940

[43] W. Humphrey, A. Dalke and K. Schulten, "VMD: visual molecular dynamics," *Journal of molecular graphics,* vol. 14, no. 1, pp. 33-38, 1996. doi: 10.1016/0263-7855(96)00018-5

[44] R. Tran, Z. Xu, B. Radhakrishnan, D. Winston, W. Sun, K. A. Persson and S. P. Ong, "Surface energies of elemental crystals," *Scientific data,* vol. 3, no. 1, pp. 1-13, 2016. doi: 10.1038/sdata.2016.80

[45] A. Miedema, "Simple Model for Alloys. Pt. 2. Influence of Ionicity on the Stability and Other Physical Properties of Alloys," *Philips Tech. Rev,* vol. 33, no. 7, pp. 196-202, 1973.

[46] A. Miedema, "Simple model for alloys," *Philips Tech. Rev,* vol. 33, no. 6, pp. 149-160, 1973.

[47] A. D. R. &. G. W. Dębski, "New features of Entall database: comparison of experimental and model formation enthalpies," *Archives of Metallurgy and Materials,* vol. 59, no. 4, pp. 1337-1343, 2014. doi: [10.2478/amm-2014-0228](10.2478/amm-2014-0228)

[48] A. V. Ruban, H. L. Skriver and J. K. Nørskov, "Surface segregation energies in transition-metal alloys," *Physical Review B,* vol. 59, no. 24, p. 15990, 1999. doi: [10.1103/PhysRevB.59.15990](10.1103/PhysRevB.59.15990)